\documentclass[aps, showkeys, nofootinbib, superscriptaddress, twocolumn]{revtex4-2}

\usepackage{graphicx}
\usepackage{epsfig}
\usepackage{bm}
\usepackage{color}
\usepackage{amsmath}
\usepackage[colorlinks, linkcolor=red,anchorcolor=green,citecolor=blue]{hyperref}

\begin{document}

\title{A numerical algorithm for solving the coupled Schr\"odinger equations\\ using inverse power method}
\author{Jiaxing Zhao}
\email{jzhao@subatech.in2p3.fr}
\address{Physics Department, Tsinghua University, Beijing 100084, China.}
\address{UBATECH, Universit\'e de Nantes, IMT Atlantique, IN2P3/CNRS, 4 rue Alfred Kastler, 44307 Nantes cedex 3, France.}
\author{Shuzhe Shi}
\email{shuzhe-shi@tsinghua.edu.cn}
\address{Physics Department, Tsinghua University, Beijing 100084, China.}

\date{\today}

\begin{abstract}
The inverse power method is a numerical algorithm to obtain the eigenvectors of a matrix. 
In this work, we develop an iteration algorithm, based on the inverse power method, to numerically solve the Schr\"odinger equation that couples an arbitrary number of components. Such an algorithm can also be applied to the multi-body systems. To show the power and accuracy of this method, we also present an example of solving the Dirac equation under the presence of an external scalar potential and a constant magnetic field, with source code publicly available .
\end{abstract}

\keywords{coupled Schr\"odinger-like equations, eigenvalue and eigenstate, inverse power method}

\maketitle

\section{Introduction}\label{sec.introduction}
In quantum physics, all properties of a microscopic state are encoded in its wavefunction. For instance, a single particle wavefunction $\psi({\bm r},t)$ represents the probability amplitude of the particle to appear at position $\bm r$ at time $t$. The properties and time evolution of the wave-function satisfy the Schr\"odinger equation. Furthermore, if the potential does not change with time --- which is the situation discussed in this paper --- one may solve the energy eigenvalues and eigenfunctions of the time-independent Schr\"odinger equation and access the complete information of the quantum system.
For a particle in a one-dimensional potential field $V(x)$, the Schr\"odinger equation is:
\begin{equation}
-\frac{\hbar^2}{2m} \frac{\mathrm{d}^2}{\mathrm{d}x^2} \psi(x) + V(x) \psi(x) = E\, \psi(x),
\label{eq.schrodinger_1d}
\end{equation}
where $-\frac{\hbar^2}{2m} \frac{\mathrm{d}^2}{\mathrm{d}x^2} \equiv \frac{\hat{p}^2}{2m}$ is the kinetic energy, $E$ is the energy, and $\hat{H}\equiv\frac{\hat{p}^2}{2m}+V$ is called the Hamiltonian. 
Physics properties impose boundary conditions for the wavefunctions. For the bound-states, one would expect that the particle's probability density vanishes when $x$ approaches infinity (or at the boundaries of a finite-size system). By imposing a boundary condition, Eq.~\eqref{eq.schrodinger_1d} becomes a Sturm--Liouville problem, and has discrete energy levels. They are referred to as the energy eigenvalues, while the corresponding wavefunctions are called the eigenfunctions.

While Eq.~\eqref{eq.schrodinger_1d} describes the quantum states for a particle moving in one dimension and has a simple form, many physics systems need to be described by Schr\"odinger equations in more complex forms. One example is a one-dimensional particle with extra discrete degrees of freedom, such as orbital angular momentum, spin, isospin, color, etc. In such systems, the Schr\"odinger equation becomes a coupled equation:
\begin{equation}
-\frac{\hbar^2}{2m} \frac{\mathrm{d}^2}{\mathrm{d}x^2} \psi_i(x) + \sum_j V_{ij}(x) \psi_i(x) = E\, \psi_i(x), \label{eq.schrodinger_2}
\end{equation}
where $i=0$, $1$, $\cdots$ labels the state for the extra discrete degree(s) of freedom, and the off-diagonal elements of the potential matrix, $V_{ij}$, describe the transition rate between different states.

Another example is a multi-body system, such as a molecule containing few atoms, an atom with multiple electrons, an atomic nucleus with multiple nucleons, or a hadron with multiple quarks.
Such systems can be described by the multi-body Schr\"odinger equation,
\begin{align}\begin{split}
&\;	\left(\sum_{i=1}^N \frac{\hat {\bf p}^2_i}{ 2m_i}+ V({\bf r}_1,{\bf r}_2,...,{\bf r}_N)\right ) \Psi({\bf r}_1,{\bf r}_2,...,{\bf r}_N) \\
=&\; E \,\Psi({\bf r}_1,{\bf r}_2,...,{\bf r}_N), \label{eq.schrodinger_3}
\end{split}\end{align}
where $\Psi({\bf r}_1,{\bf r}_2,...,{\bf r}_N)$ is the $N$-body wavefunction, $V({\bf r}_1,{\bf r}_2,...,{\bf r}_N)$ is the multi-body interaction potential.

One can convert Eq.~\eqref{eq.schrodinger_3} into the form of Eq.~\eqref{eq.schrodinger_2} by performing coordinate transformation and basis expansion. Detailed procedures can be found in e.g.,~\cite{Zhao:2020nwy, Zhao:2023qww} and we briefly review them as follows.
To simplify such an $N$-body Schr\"odinger equation, one can employ the Jacobi coordinates to transform the individual coordinates into a center-of-mass coordinate ${\bf R}$ and $N-1$ relative coordinates ${\bf x}_1, {\bf x}_2,...,{\bf x}_{N-1}$. By separating the center-of-mass motion, only the relative motion needs to be considered.
The relative coordinates can be further transformed into a hyper-radius $\rho\equiv\sqrt{x_1^2+x_2^2+...+x_{N-1}^2}$ and $3N-4$ hyper-angles $\Omega=\{\alpha_2,\alpha_3,...,\alpha_{N-1}, \theta_1,\phi_1, \theta_2,\phi_2,...,\theta_{N-1},\phi_{N-1}\}$.
The definition of hyper-angle is $\sin \alpha_i=x_i/\rho_i$ with $\rho_i=(\sum_{j=1}^i x_j^2)^{1/2}$. $(x_i, \theta_i, \phi_i)$ are the spherical coordinates corresponding to ${\bf x}_i$. 
In such hyper-spherical coordinates, one can expand the wavefunction in a series of the complete and orthonormal hyper-spherical harmonic functions ${\mathcal Y}_\kappa(\Omega)$ 
via $\Psi=\sum_\kappa R_\kappa(\rho){\mathcal Y}_\kappa(\Omega)$, where $\kappa$ is a combined index labeling all quantum numbers of the hyper-spherical harmonic functions.
The $N$-body Schr\"odinger equation becomes a coupled differential equation for different radial wavefunctions $R_\kappa(\rho)$:
\begin{align} \begin{split}
&\frac{\hbar^2}{2\mu}\left(
	- \frac{\mathrm{d}^2}{\mathrm{d}\rho^2} - \frac{3N-4}{\rho}\frac{\mathrm{d}}{\mathrm{d}\rho}
	+ \frac{K(K+3N-5)}{\rho^2}\right) R_{\kappa} \\
& +\sum_{\kappa'}V_{\kappa \kappa'}R_{\kappa'} = E \, R_{\kappa}.
\end{split}\end{align}
where $\mu$ is the reduced mass, $K$ is the hyper-angular quantum number determined by ${\mathcal Y}_\kappa$, and the potential matrix element is $V_{\kappa \kappa'} \equiv \int V(\rho, \Omega)  {\mathcal Y}_\kappa^*(\Omega) {\mathcal Y}_{\kappa'}(\Omega)\mathrm{d}\Omega$. 
One can refer to e.g.~\cite{Krivec:1998,Barnea:1999be,Marcucci:2019hml,Zhao:2020jqu,Dohet-Eraly:2019wdb} for more details about Jacobi coordinates and hyper-spherical harmonic functions. 

The Schr\"odinger equations expressed by Eqs.~(\ref{eq.schrodinger_1d}--\ref{eq.schrodinger_3}) cover a wide variety of physical systems, and solving them is a crucial problem of many topics in quantum mechanics. Here are a few examples: 
molecule and atom bound-states~\cite{anderson1975random, lester1990quantum, mitas1996electronic},
few-nucleon bound-states~\cite{Kameyama:1989zz, Hiyama:2011ge}, multi-quark bound-states (baryon/tetraquark/pentaquark)~\cite{Hiyama:2005cf, Zhao:2023qww, Zhao:2016ccp, Yang:2018oqd, Wang:2019rdo, Zhao:2020nwy}, two-body Dirac equation for mesons and electron-position~\cite{Crater:1987hm, Crater:2010fc, Shi:2013rga},
molecules and mesons under the presence of external magnetic field~\cite{Tiesinga:1993zza, Alford:2013jva, Guo:2015nsa, Chen:2020xsr},  etc.
Nevertheless, even in the simple one-dimension case~\eqref{eq.schrodinger_1d}, there is only a few interaction potentials that one can find the analytic solutions, e.g., the harmonic oscillators and the Coulomb potential. In most cases, one would have to solve the Schr\"odinger equations numerically.

A few numerical algorithms are developed to solve the eigenstate problem. Among them, the variational method, the diffusion Monte Carlo (DMC), the finite difference time domain (FDTD) method, and the inverse power method are commonly employed. The variational method is a textbook technique. It starts from a ``trial'' wavefunction, with a particular parameterization, and computes the energy as the expectation of the Hamiltonian. Then the ground state could be obtained when the energy is minimized by varying the parameters. On the other hand, both the DMC and FDTD methods share the same idea to solve the  evolution of imaginary-time wavefunction ruled by $\partial_\tau \Psi(\tau, {\bf x}) = -\hat{H} \Psi(\tau, {\bf x})$. The time dependence can be formulated as $\Psi(\tau, {\bf x}) = \sum_n c_n e^{-E_n \tau} \psi_n({\bf x})$, so that the excitations  --- the states with higher energies --- decay faster than the ground state. After a sufficiently long time, the wavefunction contains the ground state only. We refer the readers to Refs.~\cite{anderson1975random, lester1990quantum, mitas1996electronic, grimm1971monte, foulkes2001quantum} for more details of the DMC, and Refs.~\cite{feagin2002quantum, roy2001time, wadehra2003ground, sullivan2005determining, sullivan2005determining2, roy2005low, Sudiarta:2007, Strickland:2009ft} about the FDTD.
All these algorithms are designed to find the ground state --- the quantum state with the lowest energy.  One is possible to find the excitations by imposing an additional requirement that the wavefunction of interest shall be orthogonal to the known lower energy states. Hence, to solve the excitations, it is necessary to first find the lower energy states, and their wave function needs to be determined with high precision.

Apart from the aforementioned algorithms, in Ref.~\cite{1994JCoPh.115..470C} H.~W.~Crater developed a new algorithm, based on the inverse power method (IPM), to solve the Schr\"odinger equations for non-coupled one-dimension systems and the Schr\"odinger equations with two or four coupling components.
The IPM algorithm is efficient in two aspects: first, it directly searches for the quantum state including both ground state and excitations; second, it converges faster towards the eigenstate of interest.
In the current work, we generalize Crater's IPM algorithm and make it applicable to solve the Schr\"odinger equation that couples an arbitrary number of components, with the generic form:
\begin{equation}
-\frac{\hbar^2}{2m} \frac{\mathrm{d}^2\psi_i(r)}{\mathrm{d}r^2}
+ \sum_{j=1}^{M} \Big(F_{ij}(r) \frac{\mathrm{d}}{\mathrm{d}r} + U_{ij}(r)\Big) \psi_j(r) 
= E\, \psi_i(r),\label{eq.coupled_schroedinger}
\end{equation}
for $i=1$, $\cdots$, $M$, together with the boundary conditions
\begin{equation}
\psi_i (r_\mathrm{min}) = \psi_i (r_\mathrm{max}) = 0\,.
\end{equation}
Here, $r_\mathrm{min}$ ($r_\mathrm{max}$) is the left (right) boundary, and can be either finite or infinite.

The structure of this paper is as follows. For the pedagogical reason, we begin with reviewing the IPM and its application to non-coupled Schr\"odinger equations in Sec.~\ref{sec.non_coupled}. Then in Sec.~\ref{sec.coupled}, we derive the IPM algorithm for the coupled Schr\"odinger equations. Finally, in Sec.~\ref{sec.example}, we will take the Dirac equation under the presence of an external scalar potential and a constant magnetic field as an example, and show how to employ the IPM to solve the energy eigenvalues and eigenfunctions. A summary is given in Sec.~\ref{sec.summary}.

\section{Inverse Power Method for Non-Coupled Schr\"odinger-like Equations}\label{sec.non_coupled}
We begin with briefly reviewing the IPM, which is an iterative algorithm to numerically solve the eigenvectors of a matrix.
For better connection with later sections, we denote the matrix as $\hat H$, the eigenvectors as $|\psi_k\rangle$, and the corresponding eigenvalues as $E_k$. They respectively correspond to the Hamiltonian operator, eigenstates, and energy eigenvalues in an eigenstate problem in quantum physics.
We formally write down the eigenvalue problem as
\begin{equation}
\hat H \,| \psi_k \rangle = E_k \,| \psi_k \rangle \,,\qquad (k = 1, 2, \cdots),\label{eq.eigen}
\end{equation}
while the set formed by all the eigenvectors, $\{|\psi_k\rangle\}$, is a complete basis of the orthonormal vectors:
\begin{equation}
\langle \psi_k | \psi_{k'} \rangle = \delta_{kk'}\,, \qquad
\sum_k | \psi_k \rangle\langle \psi_k | = \hat I \,.
\end{equation}
In above equation, $\hat I$ is a unity operator, $\hat I |\psi\rangle = |\psi\rangle$, for arbitrary vector $|\psi\rangle$. In the linear algebra representation, one can consider the right-bracket as a column vector ($|\psi\rangle\equiv\mathbf{v}_{\psi}$), while the left-bracket is a row vector being the conjugate transpose of the right-bracket($\langle \psi|\equiv \mathbf{v}_{\psi}^\dagger$). With these, the inner product is defined as $\langle \psi_k | \psi_{k'} \rangle \equiv \mathbf{v}_{\psi_k}^\dagger \cdot \mathbf{v}_{\psi_{k'}} = \delta_{k,k'}$, while the completeness relation becomes $I = \sum_k | \psi_k \rangle\langle \psi_k | \equiv\sum_k \mathbf{v}_{\psi_{k}}  \cdot \mathbf{v}_{\psi_k}^\dagger$, where $I$ is a unity matrix.

For any arbitrary scalar value $\lambda$, one can show that $(\hat H - \lambda \hat I)^{-1}$ shares the same eigenvectors of $\hat H$,
\begin{equation}
(\hat H - \lambda \hat I)^{-1} \,| \psi_k \rangle = (E_k - \lambda)^{-1} \,| \psi_k \rangle\,,
\end{equation}
and likewise for the power of $(\hat H - \lambda \hat I)^{-1}$,
\begin{equation}
\big((\hat H - \lambda \hat I)^{-1}\big)^n \,| \psi_k \rangle = (E_k - \lambda)^{-n} \,| \psi_k \rangle\,.
\end{equation}
Based on such a relation, the IPM approximately gives the eigenvector of $\hat{H}$ with an eigenvalue closest to $\lambda$ in an iterative manner.
One may start the iteration from an arbitrary vector $|\psi^{(0)} \rangle$, which can be formally decomposed in the basis of $\{ |\psi_k\rangle\}$,
\begin{equation}
|\psi^{(0)} \rangle = \sum_k c_k^{(0)} |\psi_k\rangle.
\end{equation}
We repeatedly multiply the vector by $(\hat H - \lambda \hat I)^{-1}$ and normalized it,
\begin{align}
&|{\phi}^{(i)} \rangle \equiv 
    (\hat H - \lambda \hat I)^{-1} |\psi^{(i-1)} \rangle\,,\\
&|\psi^{(i)}\rangle \equiv 
    |{\phi}^{(i)} \rangle / \langle {\phi}^{(i)} |{\phi}^{(i)} \rangle^{\frac{1}{2}}  \,, \qquad{i=1,2,\cdots.}
\end{align}
After $n$ iterations, we obtain
\begin{align}
\begin{split}
|\psi^{(n)} \rangle =&\; N_n \, [(\hat H - \lambda \hat I)^{-1}]^n |\psi^{(0)} \rangle \\
=&\; N_n \sum_k(E_k-\lambda)^{-n} c_k^{(0)} |\psi_k\rangle,
\end{split}
\end{align}
where $N_n$ is the normalization factor ensuring $\langle \psi^{(n)} | \psi^{(n)}\rangle = 1$.
The coefficients of $|\psi^{(n)} \rangle$ in the $\{ |\psi_k\rangle\}$ basis can be written as
\begin{equation}
c_k^{(n)} \equiv \langle \psi_k | \psi^{(n)} \rangle = N_n (E_k-\lambda)^{-n} c_k^{(0)}\,.
\end{equation}
For large enough $n$, $|\psi^{(n)}\rangle$ is dominated by the eigenvector, which corresponds to the eigenvalue closest to $\lambda$. 
We label such an eigenvector as $|\psi_\ell\rangle$, and its corresponding eigenvalue, $E_\ell$, satisfies
\begin{equation}
|E_\ell - \lambda| < |E_i - \lambda|, \qquad \forall i\neq \ell.
\end{equation}
It is straightforward to show that
\begin{equation}
|E_\ell - \lambda|^{-n} \gg |E_i - \lambda|^{-n}, \qquad \forall i\neq \ell,
\end{equation}
for large enough $n$.
Supposing that the initial projection coefficients $c_k^{(0)}$'s are of the same order of magnitude, one can find that $c_\ell^{(n)} \gg c_i^{(n)}$ for all $i\neq \ell$.
After a sufficiently larger number of iterations,
\begin{equation}
\lim_{n\to\infty} \frac{|E_i - \lambda|^{-n}}{|E_\ell - \lambda|^{-n}} = 0 \,,
\end{equation}
and hence $|\psi^{(n)}\rangle$ converges to the eigenvector
\begin{equation}
|\psi_\ell\rangle = \lim_{n\to\infty} |\psi^{(n)}\rangle\,,
\end{equation}
while the corresponding eigenvalue
\begin{equation}
E_\ell = \lambda + \lim_{n\to\infty} \frac{1}{\langle \psi^{(n)}| (\hat H - \lambda \hat I)^{-1} |\psi^{(n)}\rangle}\,.
\end{equation}

Now we move on to discuss how one can employ the IPM to solve the eigenstate problem of a non-coupled Schr\"odinger-like equation:
\begin{align}\begin{split}
&\widehat H\, \psi(r) \equiv \Big(-\frac{\hbar^2}{2m} \frac{\mathrm{d}^2}{\mathrm{d}r^2} + F(r)  \frac{\mathrm{d}}{\mathrm{d}r}+ U(r)\Big) \psi(r) 
= E\, \psi(r), \\
&\psi (r_\mathrm{min}) = \psi (r_\mathrm{max}) = 0.
\end{split}\end{align}
The boundaries $r_\mathrm{min}$ and $r_\mathrm{max}$ can be infinite in physical problems.
Practical numerical computations always truncate the range of $r$ by requiring that $[r_\mathrm{min}, r_\mathrm{max}]$ should be big enough to capture the physics of interest.
A commonly used technique is to discretize the space, and convert the continuous wavefunctions into their values at coordinates $r_\mathrm{min}$, $r_\mathrm{min}+h$, $r_\mathrm{min}+2h$, $\cdots$, $r_\mathrm{min}+Nh$,
\begin{align}
\psi(r) \to &\; \left(\psi_{[0]}, \psi_{[1]}, \psi_{[2]}, \cdots, \psi_{[N]} \right)^T \,,
\end{align}
where $h$ is the grid spacing and $N \equiv (r_\mathrm{max}-r_\mathrm{min})/h$ is the number of grids. Here and for the rest of this paper, we use subscripted square brackets, $f_{[\cdot]}$, to label the discretized indices, while parentheses, $f(\cdot)$, to indicate the dependence of continuous variables. The discretized labeling is defined as $f_{[k]} \equiv f(r=k\,h)$.
Besides, we use the superscript parentheses, $f^{(\cdot)}$, to denote IPM iterations.
The boundary condition requires $\psi_{[0]} = \psi_{[N]} = 0$.
Then, the first-order derivative becomes
\begin{align}
\frac{\mathrm{d}}{\mathrm{d}r} \psi(r) \to \frac{1}{2h} 
\left(\begin{array}{cccc}
0 & 1 &  & \\
-1 & \ddots & \ddots & \\
& \ddots & \ddots & 1\\
 & & -1 & 0 \\
\end{array}\right) \cdot
\left(\begin{array}{c}
\psi_{[0]} \\ \vdots\\ \vdots \\ \psi_{[N]}
\end{array}\right) \,,
\end{align}
the second-order derivative reads
\begin{align}
\frac{\mathrm{d}^2}{\mathrm{d}r^2} \psi(r) \to \frac{1}{h^2} 
\left(\begin{array}{cccc}
-2 & 1 &  & \\
1 & \ddots & \ddots & \\
& \ddots & \ddots & 1\\
 & & 1 & -2 \\
\end{array}\right) \cdot
\left(\begin{array}{c}
\psi_{[0]} \\ \vdots\\ \vdots \\ \psi_{[N]}
\end{array}\right) \,,
\end{align}
and the potential term
\begin{align}
U(r) \psi(r) \to 
\left(\begin{array}{ccc}
U_{[0]} & & \\
 & \ddots & \\
 &  & U_{[N]} \\
\end{array}\right) \cdot
\left(\begin{array}{c}
\psi_{[0]} \\ \vdots \\ \psi_{[N]}
\end{array}\right) \,.
\end{align}

Discretizing the spatial coordinate allows one to convert the eigenvalue problem of a differential equation into that of a matrix, but the application of IPM is not straightforward. The main challenge comes from computing the inverse matrix $(\hat H - \lambda)^{-1}$, where the Hamiltonian operation matrix is an $N \times N$-dimensional large matrix.
Such difficulty has been solved by H.~W.~Crater in Ref.~\cite{1994JCoPh.115..470C}, and in the rest of this section we review his method, and then in the next section we generalize the technique to the coupled Schr\"odinger-like equations~\eqref{eq.coupled_schroedinger}.

Following the procedures of IPM, one can begin with a trial eigenvalue $\lambda$ and a trial eigenstate $\psi^{(0)}$, and define the function series as
\begin{align}\begin{split}
\phi^{(l+1)} =&\; (\hat H - \lambda)^{-1} \psi^{(l)}, \\
\psi^{(l+1)} =&\; \phi^{(l+1)} / \langle \phi^{(l+1)} |  \phi^{(l+1)} \rangle^{\frac{1}{2}}.
\end{split}\label{eq.nc_ipm_0}
\end{align}
where $\langle \phi^{(l+1)} |  \phi^{(l+1)} \rangle \equiv h \sum_{k=1}^{N} |\phi^{(l+1)}_{[k]}|^2$.

Instead of solving $ (\hat H - \lambda)^{-1}$ explicitly, Ref.~\cite{1994JCoPh.115..470C} rewrite the iteration equation~\eqref{eq.nc_ipm_0} as 
\begin{equation}
 (\hat H - \lambda) \phi^{(l+1)} = \psi^{(l)} \label{eq.nc_ipm_1}
\end{equation}
and solve the unknown $\phi^{(l+1)}$ from the known $\psi^{(l)}$.
For simplification, we define,
\begin{align}
c_{[k]} \equiv&\; - \frac{\hbar^2}{2mh^2} - \frac{F_{[k]}}{2h} \,,\\
d_{[k]} \equiv&\; U_{[k]}-\lambda + \frac{\hbar^2}{mh^2}  \,,\\
e_{[k]} \equiv&\; - \frac{\hbar^2}{2mh^2} + \frac{F_{[k]}}{2h} \,,
\end{align}
and \eqref{eq.nc_ipm_1} can be written explicitly as
\begin{align}
\psi^{(l)}_{[k]} = &\;
	c_{[k]} \phi^{(l+1)}_{[k-1]}
	+ d_{[k]} \phi^{(l+1)}_{[k]}
	+ e_{[k]} \phi^{(l+1)}_{[k+1]} \,,
 \label{eq.nc_ipm_2}
\end{align}
for $1 \leq k \leq N-1$, while on the boundary, 
the bound-state requirement indicates that
\begin{equation}
\psi^{(l)}_{[0]} = \psi^{(l)}_{[0]} =
\phi^{(l)}_{[N]} =\phi^{(l)}_{[N]}=0,
\end{equation}
hence the finite differences at boundaries remain well-defined.

To solve $\phi^{(l+1)}$, one can introduce the auxiliary vectors
\begin{align}
\overline d_{[1]} \equiv&\; d_{[1]}, \\
\overline \psi_{[1]} \equiv&\; \psi^{(l)}_{[1]} ,
\end{align}
and
\begin{align}
\overline d_{[k]} \equiv&\; d_{[k]} - c_{[k]} e_{[k-1]} / \overline d_{[k-1]}\,,\\
\overline \psi_{[k]} \equiv&\; \psi^{(l)}_{[k]} - c_{[k]} \overline\psi_{[k-1]} / \overline d_{[k-1]}\,,
\end{align}
for $k = 2, 3, \cdots, N$.  Then one can take the substitutions
\begin{align}
d_{[k]} =&\; \overline d_{[k]} + c_{[k]} e_{[k-1]} / \overline d_{[k-1]} \,, \\
\psi^{(l)}_{[k]} =&\; \overline \psi_{[k]} + c_{[k]} \overline\psi_{[k-1]} / \overline d_{[k-1]} \,,
\end{align}
in \eqref{eq.nc_ipm_2}, and rewrite it as
\begin{equation}
\overline d_{[1]} \phi^{(l+1)}_{[1]} + e_{[1]} \phi^{(l+1)}_{[2]}  - \overline\psi_{[1]} = 0 \,, \label{eq.nc_ipm_3}
\end{equation}
and
\begin{align}\begin{split}
0= &\;
	\frac{c_{[k]}}{\overline d_{[k-1]}} \Big(\overline d_{[k-1]} \phi^{(l+1)}_{[k-1]}
	+ e_{[k-1]}) \phi^{(l+1)}_{[k]}  - \overline\psi_{[k-1]} \Big)
\\&\; 
	+ \Big(\overline d_{[k]} \phi^{(l+1)}_{[k]}
	+ e_{[k]} \phi^{(l+1)}_{[k+1]} - \overline\psi_{[k]} \Big) \,,
\end{split} \label{eq.nc_ipm_4}
\end{align}
for $k>1$. Then it is clear that
\begin{equation}
\overline d_{[k]} \phi^{(l+1)}_{[k]} + e_{[k]} \phi^{(l+1)}_{[k+1]} - \overline \psi_{[k]} = 0, 
\end{equation}
for all $1\leq k \leq N-1$. 
With these, one can start from the boundary condition $\phi^{(l+1)}_{[N]} = 0$, and solve the unknown vector step-by-step
\begin{equation}
 \phi^{(l+1)}_{[k]} = \Big(\overline \psi_{[k]} - e_{[k]} \phi^{(l+1)}_{[k+1]} \Big) / \overline d_{[k]} \label{eq.nc_ipm_5}
\end{equation}
for $k=N-1$, $N-2$, $\cdots$, $1$.
With the solution of $\phi^{(l+1)}$ found in \eqref{eq.nc_ipm_5}, one can perform the inverse power iterations in \eqref{eq.nc_ipm_0} for such Schr\"odinger-like equations, and obtain the eigenstate with eigenvalue nearest to $\lambda$.

\section{Inverse Power Method for Coupled Schr\"odinger-like Equation}\label{sec.coupled}

In this section, we move on to discuss a more general form of Schr\"odinger-like equation, which contains other discrete degrees of freedom in addition to the coordinate $r$. 
In such case, the equation becomes a coupled equation of these extra degrees of freedom:
\begin{align}\begin{split}
&-\frac{\hbar^2}{2m} \frac{\mathrm{d}^2\psi_i(r)}{\mathrm{d}r^2}
+ \sum_{j=1}^{M} \Big(F_{ij}(r) \frac{\mathrm{d}}{\mathrm{d}r} + U_{ij}(r)\Big) \psi_j(r) 
= E\, \psi_i(r), \\
&\psi_i (r_\mathrm{min}) = \psi_i (r_\mathrm{max}) = 0 \,,
\end{split}\end{align}
for $i=1$, $2$, $\cdots$, $M$.
It shall be mentioned that in Ref.~\cite{1994JCoPh.115..470C}, H. W. Crater already obtained special solutions for the cases $M=2$ and $M=4$. 
In this section, however, we discuss a general algorithm independent of the exact value of $M$.
To get a compact form, we denote the $M$-dimensional wavefunction vector as
\begin{equation}
\mathbf \Psi_{[k]} \equiv \Big(\psi_{1}(r_k),\psi_{2}(r_k), \cdots ,\psi_{M}(r_k) \Big)^T\,,
\end{equation}
where $r_k\equiv r_\mathrm{min} + k\,h$.
The $M\times M$-dimensional matrices $\mathbf F_{[k]}$ and $\mathbf U_{[k]}$ is defined in such a way that the $i$-th row, $j$-th column of them are $F_{ij}(kh)$ and $U_{ij}(kh)$, respectively. In this section, we adopt the bold symbols to represent $M\times M$-dimensional matrices and $M$-dimensional vectors.

The inverse power procedure starts with a trial eigenvalue $\lambda$ and a trial state $\mathbf\Psi^{(0)}$, and take the iteration
\begin{align}
\mathbf\Phi^{(l+1)} =&\; (\hat{\mathbf H} - \lambda\, \mathbf I_{N M})^{-1} \cdot \mathbf\Psi^{(l)} 
\label{eq.c_ipm_0_1}\\
\mathbf\Psi^{(l+1)} =&\; \mathbf\Phi^{(l+1)} / \langle \mathbf\Phi^{(l+1)} | \mathbf\Phi^{(l+1)}\rangle^{\frac{1}{2}}
\label{eq.c_ipm_0_2}
\end{align}
where $\mathbf I_{N M}$ is an $N\,M\times N\,M$-dimensional identity matrix. The inner product is defined as $$\langle \mathbf\Phi^{(l+1)} | \mathbf\Phi^{(l+1)}\rangle \equiv h \sum_{i=1}^{M} \sum_{k=1}^{N} |\phi_{i,[k]}^{(l+1)}|^2.$$
They satisfy the boundary condition
\begin{equation}
\mathbf\Phi^{(l+1)}_{[0]} = \mathbf\Psi^{(l)}_{[0]} =
\mathbf\Phi^{(l+1)}_{[N]} = \mathbf\Psi^{(l)}_{[N]} = \mathbf0.
\end{equation}

Analogous to \eqref{eq.nc_ipm_1},
we obtain $\mathbf\Phi^{(l+1)}$ by solving 
\begin{equation}
(\hat{\mathbf H} - \lambda\,\mathbf I_{N M}) \cdot \mathbf\Phi^{(l+1)} = \mathbf\Psi^{(l)} \,,\label{eq.c_ipm_1}
\end{equation}
which has the explicit form that
\begin{align}
 \mathbf\Psi^{(l)}_{[k]} =&\; 
	 \mathbf C_{[k]} \cdot \mathbf\Phi^{(l+1)}_{[k-1]} 
	+ \mathbf D_{[k]} \cdot \mathbf\Phi^{(l+1)}_{[k]}
	+ \mathbf E_{[k]} \cdot \mathbf\Phi^{(l+1)}_{[k+1]}, 
\label{eq.c_ipm_2}
\end{align}
with 
\begin{align}
\mathbf C_{[k]} \equiv&\; - \frac{\hbar^2}{2mh^2} \mathbf I_{M} - \frac{1}{2h}\mathbf F_{[k]} \,,\label{eq:C}\\
\mathbf D_{[k]} \equiv&\; \mathbf U_{[k]}-\lambda\,\mathbf I_{M} + \frac{\hbar^2}{mh^2} \mathbf I_{M} \,,\label{eq:D}\\
\mathbf E_{[k]} \equiv&\; - \frac{\hbar^2}{2mh^2} \mathbf I_{M}+ \frac{1}{2h}\mathbf F_{[k]} \,.\label{eq:E}
\end{align}
Similar to the non-coupled case, one shall keep in mind the vanishing wavefunctions when computing derivatives at the boundary.
We define the auxiliary matrix $\overline {\mathbf D}$ and vectors $\overline {\mathbf \Psi}$, with initialization
\begin{align}
\overline {\mathbf D}_{[1]} = \mathbf D_{[1]} ,\qquad
\overline {\mathbf \Psi}_{[1]} = \mathbf \Psi^{(l)}_{[1]},
\end{align}
and then compute
\begin{align}
\overline{\mathbf D}_{[k]} =&\; \mathbf D_{[k]} - \mathbf C_{[k]} \cdot \overline{\mathbf D}_{[k-1]}^{-1} \cdot \mathbf E_{[k-1]} , \label{eq:Dbar}\\
\overline{\mathbf \Psi}_{[k]} =&\; \mathbf \Psi^{(l)}_{[k]} - \mathbf C_{[k]} \cdot \overline{\mathbf D}_{[k-1]}^{-1} \cdot \overline{\mathbf \Psi}_{[k-1]}, \label{eq:Psibar}
\end{align}
for $k=2$, $3$, $\cdots$, $N$. 
Particularly, $ \overline{\mathbf D}_{[k]}^{-1}$ stands for the inverse matrix of $ \overline{\mathbf D}_{[k]}$. Noting that both of them are $M\times M$-dimensional matrices and $M$ is typically not very large, computing the inverse is not a problem.
Similar to \eqref{eq.nc_ipm_3} and \eqref{eq.nc_ipm_4}, we can re-write \eqref{eq.c_ipm_2} as
\begin{equation}
\mathbf0 = \overline{\mathbf D}_{[1]}\cdot \mathbf \Phi^{(l+1)}_{[1]}
	+ \mathbf E_{[1]}\cdot\mathbf \Phi^{(l+1)}_{[2]} 
	- \overline{\mathbf \Psi}_{[1]} \,, \label{eq.c_ipm_3}
\end{equation}
and
\begin{align}\begin{split}
&\mathbf C_{[k]} \cdot \overline{\mathbf D}_{[k-1]}^{-1} \cdot 
	\Big( \overline{\mathbf D}_{[k-1]} \cdot \mathbf\Phi^{(l+1)}_{[k-1]} 
	+ \mathbf E_{[k-1]} \cdot \mathbf\Phi^{(l+1)}_{[k]} -\overline {\mathbf\Psi}_{[k-1]}\Big) \\
&+\Big(\overline{\mathbf D}_{[k]} \cdot \mathbf\Phi^{(l+1)}_{[k]} 
	+\mathbf E_{[k]}  \cdot \mathbf\Phi^{(l+1)}_{[k+1]} - \overline{\mathbf \Psi}_{[k]}\Big) = {\mathbf0}\,,
\end{split} \label{eq.c_ipm_4}
\end{align}
for $1 < k \leq N-1$.
Consequently, we find the vanishing list
\begin{equation}
\mathbf0 = \overline{\mathbf D}_{[k]} \cdot \mathbf\Phi^{(l+1)}_{[k]} 
	+\mathbf E_{[k]}  \cdot \mathbf\Phi^{(l+1)}_{[k+1]} - \overline{\mathbf \Psi}_{[k]},\qquad \forall k
\end{equation}
and the solution of $\mathbf \Phi^{(l+1)}$ is obtained as
\begin{equation}
\mathbf \Phi^{(l+1)}_{[k]} = \overline{\mathbf D}_{[k]}^{-1} \cdot \overline{\mathbf\Psi}_{[k]} - \overline{\mathbf D}_{[k]}^{-1} \cdot \mathbf E_{[k]} \cdot \mathbf\Phi^{(l+1)}_{[k+1]} \,.
\label{eq:Phi}
\end{equation}

With sufficient steps of iterations~\eqref{eq.c_ipm_0_1} and~\eqref{eq.c_ipm_0_2}, we can find a good approximation of the eigenstate with eigenvalue near $\lambda$,
\begin{align}
\Psi_\ell \approx&\; \Psi^{(n)} \,,\label{eq:wf_final}\\
E_\ell \approx&\; \lambda + \frac{1}{\langle \Psi^{(n-1)} | \Phi^{(n)} \rangle}\,.\label{eq:e_final}
\end{align}
We summarize the IPM steps in TABLE.~\ref{tab:iteration_procedure}.
\begin{table}[!ht]
\centering
\caption{Inverse power method iteration}
\begin{tabular}{cp{0.9\linewidth}}
\hline
\hline
{I.}& Preparation:\\
& \textit{a.} compute $\mathbf C_{[k]}$, $\mathbf D_{[k]}$, and $\mathbf E_{[k]}$ according to (\ref{eq:C}--\ref{eq:E});\\
& \textit{b.} compute $\overline{\mathbf D}_{[k]}$~\eqref{eq:Dbar} and their inverses;\\
& \textit{c.} choose an arbitrary initial wavefunction $\Psi^{(0)}_{[k]}$ and test energy $\lambda$;\\
{II.} & in the $l$-th IPM iteration:\\
& \textit{d.} compute $\overline{\mathbf \Psi}_{[k]}$~\eqref{eq:Psibar} with increasing $k$ from $1$ to $N$;\\
& \textit{e.} compute $\mathbf \Phi_{[k]}^{(l+1)}$~\eqref{eq:Phi} with decreasing $k$ from $N$ to $1$;\\
& \textit{f.} renormalize according to \eqref{eq.c_ipm_0_2} and compute $\mathbf \Psi_{[k]}^{(l+1)}$;\\
{III.} & Repeat (\textit{d}-\textit{f}) for $l = 0, 1, \cdots$, until the wavefunctions~\eqref{eq:wf_final} and energy~\eqref{eq:e_final} converge.\\
\hline
\end{tabular}
\label{tab:iteration_procedure}
\end{table}

It would be useful to estimate the numerical error of these quantities.
We denote the energy eigenvalues as $E_i$ and the eigenfunctions as $\Psi_i$, and define the ratio 
\begin{equation}
R_i \equiv \frac{|E_\ell - \lambda|}{|E_i - \lambda|} \,,
\end{equation}
which satisfies $R_\ell = 1$ and $R_{i\neq \ell}<1$ by definition.
After $n$ iteration steps, we can estimate the $\Psi_i$ component of $\Psi^{(n)}$ as
\begin{equation}
|\langle \Psi_i | \Psi^{(n)}\rangle| \sim (R_i)^n\,.
\end{equation}
Meanwhile, the error of energy
\begin{align}
\Delta E_\ell \equiv &\;	\lambda + \frac{1}{\langle \Psi^{(n-1)} | \Phi^{(n)} \rangle} - E_\ell \\
 \sim&\;	\lambda + \Big[ \sum_i (E_i - \lambda)^{-1} (R_i)^{2n} \Big]^{-1} - E_\ell \\
 \sim&\;	(E_\ell - \lambda)\sum_{\forall E_i\neq E_\ell} (R_i)^{2n+1} \,.
\end{align}
Both errors decay in powers of $R_i$, and the inverse power iteration converges faster when $\lambda$ is closer to $E_\ell$.
 
It becomes subtle when there exists degenerated states.
While the obtained eigenvalue remains unique, there is no guarantee that the inverse power iteration converges to one of the corresponding eigenstates. In practice, however, one starts with different trial initial states $\{\mathbf\Psi^{(0)}\}$ but with the same trial energy $\lambda$. After sufficient iterations, the final states $\{\mathbf\Psi^{(n)}\}$ correspond to different combinations of the degenerated states.
An orthonormal basis can be obtained after taking the Gram–Schmidt process.

Last but not least, it shall be worth noting that the IPM works not only for real matrices but also for complex ones. Consequently, the above algorithm remains valid when interaction potential $\mathbf U(r)$ contains a non-vanishing imaginary part, corresponding to the decay effect.

\section{Example: Dirac Equation under the Presence of External Potential}
\label{sec.example}
\begin{figure*}[!hbt]
    \centering
    \includegraphics[width=0.45\textwidth]{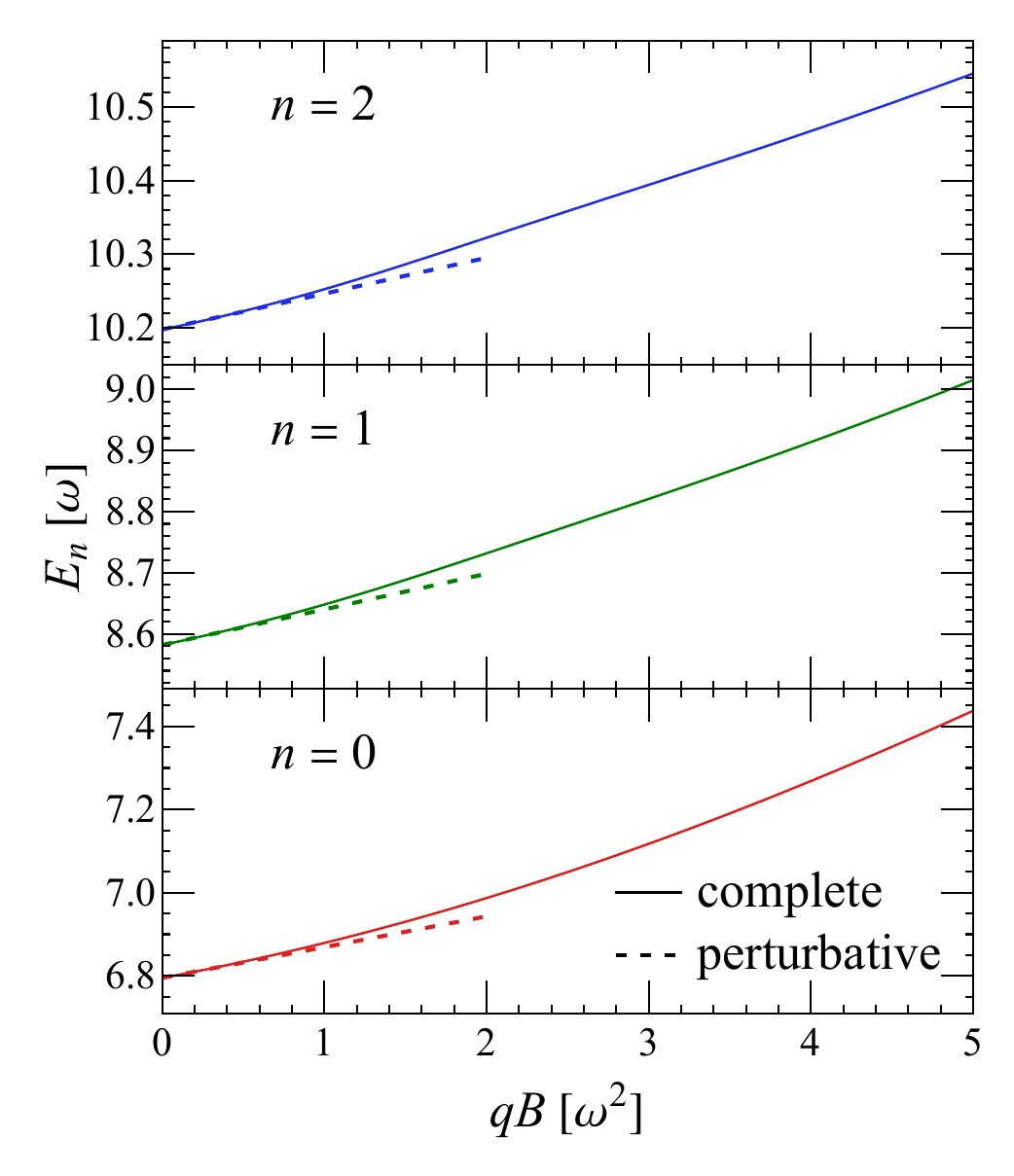}
    \includegraphics[width=0.45\textwidth]{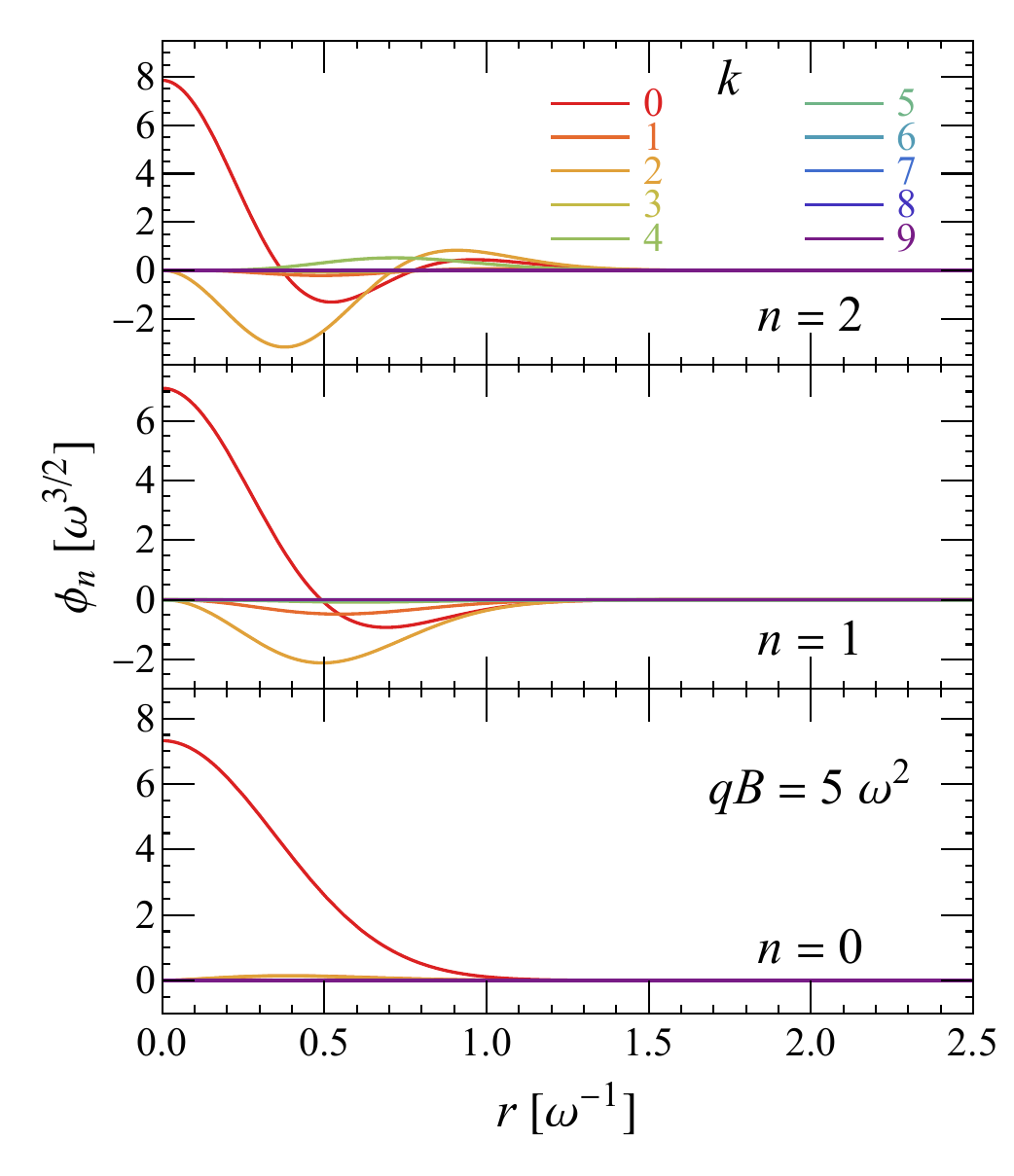}
\caption{The magnetic field-dependent energy levels(left) and wavefunctions(right) with the harmonic oscillator potential. $n$ labels the energy level. Solid lines are complete results solved by the inverse power method while dashed lines are the perturbation results from Eq.~\eqref{eq.perturbative}. The right panel corresponds to $qB=5\omega^2$, where curves with $k > 4$ are approximately zero and overlap with each other.}
    \label{fig1}
\end{figure*}
With the algorithm established, we take the Dirac equation under the presence of an external scalar central potential and a constant magnetic field as a specific physical system as an example to show how one can apply the IPM algorithm to solve the coupled Schr\"odinger-like equation. The wavefunction follows that, 
\begin{align}
    \big(\gamma^0\boldsymbol{\gamma} \cdot (-i\boldsymbol{\nabla} - q\boldsymbol{A}) + m\gamma^0 + V(r)\big) \psi(\boldsymbol{r}) = E\, \psi(\boldsymbol{r})\,,
\label{eq:Dirac}\end{align}
where $m$ is the particle mass, $q$ the electric charge, and $\boldsymbol{A} = \frac{\boldsymbol{B}\times\boldsymbol{r}}{2}$ the gauge potential, with $\boldsymbol{B} = B \hat{z}$ being the homogeneous constant magnetic field along $z$-axis. We have taken the natural unit that $\hbar = c = 1$. We take the convention of the Dirac matrices as $\gamma^0 = \Big(\begin{array}{cc}
I &  \\ & -I
\end{array}\Big)$ and $\gamma^i = \Big(\begin{array}{cc}
 & \sigma^i \\ -\sigma^i & 
\end{array}\Big)$ with $\sigma^i$ being the Pauli matrices and $I$ the two-by-two identity matrix. One may express the four-component wavefunction ($\psi$) by the two-component ones ($\phi$ and $\chi$), $\psi = \Big(\begin{array}{c}\phi \\ \chi\end{array}\Big)$.
The Dirac equation~\eqref{eq:Dirac} becomes 
\begin{align}
\begin{split}
    \big(\boldsymbol{\sigma} \cdot (-i\boldsymbol{\nabla} - q\boldsymbol{A})\big)\,\chi =\,& (E-m-V)\,\phi\,,\\
    \big(\boldsymbol{\sigma} \cdot (-i\boldsymbol{\nabla} - q\boldsymbol{A})\big)\,\phi =\,& (E+m-V)\,\chi\,,
\label{eq.2wf}    
\end{split}
\end{align}
which leads to the second-order coupled equations
\begin{align}
\begin{split}
&   r\,\Big(-\boldsymbol{\nabla}^2 
    + \frac{q^2}{4}    (\boldsymbol{B}\times\boldsymbol{r})^2
    - q\, \boldsymbol{B} \cdot (\boldsymbol{L}+\boldsymbol{\sigma})
\\&
    +\frac{V'}{r} \frac{\boldsymbol{\sigma} \cdot \boldsymbol{L}-\boldsymbol{r \cdot \nabla}}{E+m-V} 
    -\frac{q\,V'}{2r}
    \frac{(\boldsymbol{\sigma}\cdot \boldsymbol{B}) r^2 - (\boldsymbol{\sigma}\cdot\boldsymbol{r})(\boldsymbol{B}\cdot\boldsymbol{r})}{E+m-V}
    \Big)\phi 
\\=\;&   
    \big((E-V)^2-m^2 \big)r\, \phi\,.
\end{split}\label{eq:Schroedinger}
\end{align}
We have multiplied $r$ at both sides of the equality for later convenience.
The energy appears on both sides of Eq.~\eqref{eq:Schroedinger} and shall be solved in a self-consistent manner.
With $\phi$, one can also directly obtain the other wavefunction component $\chi$ via Eq.~\eqref{eq.2wf}.
We define the total angular momentum as $\boldsymbol{J} = \frac{\boldsymbol{\sigma}}{2} + \boldsymbol{L}$.
In the absence of magnetic field, it is not hard to find that $J_z$ and the magnitudes of the total angular momentum ($\boldsymbol{J}^2$), orbital angular momentum ($\boldsymbol{L}^2$), and spin ($\frac{\boldsymbol{\sigma}^2}{4}$) are all conserved.
Noting that spin can only be one-half, we therefore take the $|j,j_z,l\rangle$ basis to represent the angular momentum states, and we further take the shorthand that $|j,j_z\rangle_\pm \equiv |j,j_z,l=j\pm\frac{1}{2}\rangle$. Their explicit formulae are
\begin{align}  
|j,j_z\rangle_+ =\;&
    \left(\begin{array}{c}
    -\sqrt{\frac{j+1-j_z}{2(j+1)}} Y_{j+\frac{1}{2},j_z-\frac{1}{2}}(\theta,\varphi)    \\
    \sqrt{\frac{j+1+j_z}{2(j+1)}} Y_{j+\frac{1}{2},j_z+\frac{1}{2}}(\theta,\varphi)
    \end{array}\right)\,,\\
|j,j_z\rangle_- =\;&
     \left(\begin{array}{c}
    \sqrt{\frac{j+j_z}{2j}} Y_{j-\frac{1}{2},j_z-\frac{1}{2}}(\theta,\varphi)    \\
    \sqrt{\frac{j-j_z}{2j}} Y_{j-\frac{1}{2},j_z+\frac{1}{2}}(\theta,\varphi)
    \end{array}\right)\,.
\end{align}
We may expand the wavefunction as
\begin{align}
\begin{split}
    \phi(\boldsymbol{r}) =\;\;& 
    \frac{1}{r}\sum_{j,j_z}
    \Big(\phi^{+}_{j,j_z}(r) \,|j,j_z\rangle_+
    + \phi^{-}_{j,j_z}(r) \,|j,j_z\rangle_-\Big),
\end{split}
\end{align}
and Eq.~\eqref{eq:Schroedinger} becomes a Schr\"odinger-like equation with different angular momentum terms coupled with each other. The diagonal part reads
\begin{align}
\begin{split}
    &   r\,\Big(-\boldsymbol{\nabla}^2 
    - q\, \boldsymbol{B} \cdot (\boldsymbol{L}+\frac{\boldsymbol{\sigma}}{2})
    +\frac{V'}{r} \frac{\boldsymbol{\sigma} \cdot \boldsymbol{L}-\boldsymbol{r \cdot \nabla}}{E+m-V}
    \Big)\phi
\\=&
    \sum_{j,j_z,\pm}\Big(-\frac{\mathrm{d}^2}{\mathrm{d}r^2}
    + \frac{(j\pm\frac{1}{2})(j+1\pm\frac{1}{2})}{r^2}
    - q\,B\,j_z
\\&
    - \frac{V'}{E+m-V} \frac{\mathrm{d}}{\mathrm{d}r}
    \mp \frac{V'}{r} \frac{j+\frac{1}{2}}{E+m-V} 
    \Big) \phi^{\pm}_{j,j_z}(r) \,|j,j_z\rangle_\pm\,,
\end{split}
\label{eq:ham_1}
\end{align}
whereas the gyro term couples different angular momentum states
\begin{align}
\begin{split}
&   
    \frac{q^2}{4}(\boldsymbol{B}\times\boldsymbol{r})^2|j, j_z\rangle_\pm
\\=\;&
    \frac{q^2B^2r^2}{4}\Big(
    c_{j,j_z}|j, j_z\rangle_\pm
\\&
    +b_{j,j_z} |j-1, j_z\rangle_\mp
    +b_{j+1,j_z} |j+1, j_z\rangle_\mp
\\&
    - a_{j}^{j_z} a_{j-1}^{j_z} |j-2, j_z\rangle_\pm
    - a_{j+2}^{j_z} a_{j+1}^{j_z} |j+2, j_z\rangle_\pm
    \Big)\,.
\end{split}
\label{eq:ham_2}
\end{align}
We have defined notations $a_{j}^{j_z} \equiv \frac{\sqrt{j^2-j_z^2}}{2j}$, and $b_{j,j_z} \equiv \frac{j_z\,a_{j}^{j_z}}{j^2-1} $, and $c_{j,j_z} \equiv \frac{1}{2}\big(1+\frac{j_z^2}{j(j+1)}\big)$.
One can also expand the spin-magnetic-dipole coupling
\begin{align}
\begin{split}
&   
    \frac{q}{2} \Big(\frac{r\,V'}{E+m-V}-1\Big)(\boldsymbol{\sigma}\cdot \boldsymbol{B})
    | j, j_z\rangle_\pm
\\=\;&
    \frac{q\,B}{2j+(1\pm1)} \Big(\frac{r\,V'}{E+m-V}-1\Big)
\\&\times
    \Big(\mp j_z | j, j_z\rangle_\pm
    - \sqrt{(j+\frac{1\pm1}{2})^2-j_z^2} \, 
    | j\pm1, j_z\rangle_\mp\Big)\,,
\end{split}
\label{eq:ham_3}
\end{align}
and the spin-magnetic-tensor coupling
\begin{align}
\begin{split}
&   
    \frac{q\,V'}{2r}
    \frac{(\boldsymbol{\sigma}\cdot\boldsymbol{r})(\boldsymbol{B}\cdot\boldsymbol{r})}{E+m-V}
    | j, j_z\rangle_\pm
\\=\;&   
    \frac{q\,B\,r\,V'}{2(E+m-V)}
    \Big( \frac{j_z}{2j(j+1)} | j, j_z\rangle_\pm 
\\&
    - a_{j}^{j_z} | j-1, j_z\rangle_\mp
    - a_{j+1}^{j_z} | j+1, j_z\rangle_\mp\Big)\,.
\end{split}
\label{eq:ham_4}
\end{align}

With different terms in the Hamiltonian represented by Eqs.~(\ref{eq:ham_1}--\ref{eq:ham_4}), we note that $j_z$ is always conserved and the $|j, j_z\rangle_-$ component only couples to $|j\pm1, j_z\rangle_+$, $|j\pm2, j_z\rangle_-$, and itself. Likewise for $|j, j_z\rangle_+$.
We may further simplify the construction of the wavefunction by keeping only the components that couples with each other. For instance, a Parity even state with angular momentum along $z$ being $j_z$ takes the simple form
\begin{align}
\begin{split}
    \phi_{j_z}^{+}(\boldsymbol{r}) =\;\;& 
    \frac{1}{r}\sum_{k=0}^{\infty}
    \Big(\phi^{-}_{\frac{1}{2}+2k,j_z}(r) \,|\frac{1}{2}+2k,j_z\rangle_-
\\&\qquad
    + \phi^{+}_{\frac{3}{2}+2k,j_z}(r) \,|\frac{3}{2}+2k,j_z\rangle_+\Big).
\end{split}
\end{align}
To demonstrate the method, we use a harmonic oscillator $V = \frac{m\,\omega^2}{2} r^2$ with harmonic frequency $\omega = m/5$.
We make a truncation of the angular momentum states and keep terms with $k < 10$ and discrete the radial coordinate into $N=2500$ sites with spacing $h=10^{-3}/\omega$. 
We apply the numerical technique devised in this work to numerically solve the energy levels for the ground state, first, and second excitations as a function of magnetic field strength.
Results are shown in Fig.~\ref{fig1}, along with perturbative results assuming a small magnetic field for comparison. Perturbative results of the energies are obtained by solving the non-coupled equation~\eqref{eq:ham_1} with $B=0$ and get the zeroth order solutions, denoted as $E^{(0)}$ and $\phi_{j,j_z}^{(0),\pm}(r)$, which yields that
\begin{align}
\begin{split}
    E =\;& E^{(0)} + q\,B\,j_z \frac{\frac{2\mp(2j+1)}{4j(j+1)}I_1 - \frac{2j+1}{2j+(1\pm1)}
    }{I_2+I_3}
    + \mathcal{O}(q^2B^2),
\label{eq.perturbative}    
\end{split}
\end{align}
where 
\begin{align}
    I_1 =\;& \int \frac{r\,V' |\phi_{j,j_z}^{(0),\pm}(r)|^2}{E^{(0)}+m-V} d^3r \,,\\
    I_2 =\;& 2\int (E^{(0)}-V)|\phi_{j,j_z}^{(0),\pm}(r)|^2 d^3r \,,\\
    I_3 =\;& -\int \frac{V' \phi_{j,j_z}^{(0),\pm,*}(r) ( \frac{\mathrm{d}}{\mathrm{d}r}
    \pm \frac{j+\frac{1}{2}}{r}) \phi_{j,j_z}^{(0),\pm}(r)}{(E^{(0)}+m-V)^2} d^3r\,.
\end{align}
Good consistency is obtained at the small magnetic field, which verifies the complete numerical results.
The wavefunctions of different states $\phi_n$ are shown in the right panel of Fig.~\ref{fig1}. 
We observe that $0 < k\leq4$ states become sizable with a large magnetic field ($qB=5\,\omega^2$)\footnote{In contrast, $k=0$ is the only non-vanishing component when $qB=0$.}. Meanwhile, wavefunctions are approximately zero for $k>4$, which shows that with the truncation $M=10$ we have included a sufficient number of states.
The \texttt{c++} source code for this example is available at a \texttt{github} repository~\cite{code}.
On average, it takes $3$ iteration steps to obtain an energy eigenvalue with a precision of $10^{-8}\omega$. Each iteration takes $\approx 0.02$ second on an \texttt{Apple M2 Max} processor.

\section{Summary}
\label{sec.summary}
The Schr\"odinger equation is fundamental in quantum mechanics. It describes the energy eigenvalues and eigenstates for an interacting system.
In this work, we develop the inverse power method algorithm to solve the eigenstate problem for coupled Schr\"odinger-like equations. This is an efficient algorithm and applies to a wide variety of problems in quantum physics.

\vspace{0.5in}
\textbf{Acknowledgement ---} The authors thank Dr. Pengfei Zhuang for very insightful discussions. This work is supported by the Tsinghua University under grant No. 53330500923, NSFC under grant No. 12075129, and European Union’s Horizon 2020 research and the innovation program under grant agreement No. 824093 (STRONG-2020).


\begin{thebibliography}{40}
\bibitem{Zhao:2023qww}
J.~Zhao and S.~Shi,
Phys. Rev. C \textbf{109}, no.2, 024901 (2024).

\bibitem{Zhao:2020nwy}
J.~Zhao, S.~Shi and P.~Zhuang,
Phys. Rev. D \textbf{102}, no.11, 114001 (2020).


\bibitem{Krivec:1998}
Krivec, R., Few-Body Systems 25, 199–238 (1998).

\bibitem{Barnea:1999be}
N.~Barnea, W.~Leidemann and G.~Orlandini,
Phys. Rev. C \textbf{61} (2000), 054001.
\bibitem{Marcucci:2019hml}
L.~E.~Marcucci, J.~Dohet-Eraly, L.~Girlanda, A.~Gnech, A.~Kievsky and M.~Viviani,
Front. in Phys. \textbf{8}, 69 (2020).
\bibitem{Zhao:2020jqu}
J.~Zhao, K.~Zhou, S.~Chen and P.~Zhuang,
Prog. Part. Nucl. Phys. \textbf{114}, 103801 (2020).
\bibitem{Dohet-Eraly:2019wdb}
J.~Dohet-Eraly and M.~Viviani,
Comput. Phys. Commun. \textbf{253}, 107183 (2020).

\bibitem{anderson1975random}
J. Anderson, J. Chem. Phys. {\bf 63}, 1499 (1975).

\bibitem{lester1990quantum}
W.~A. Lester Jr., B.~L. Hammond, Ann. Rev. Phys. Chem., {\bf 41}, 283 (1990).

\bibitem{mitas1996electronic}
L. Mitas, Comp. Phys. Commun. {\bf 97}, 107 (1996).

\bibitem{Kameyama:1989zz}
H.~Kameyama, M.~Kamimura and Y.~Fukushima,
Phys. Rev. C \textbf{40}, 974-987 (1989).

\bibitem{Hiyama:2011ge}
E.~Hiyama and M.~Kamimura,
Phys. Rev. A \textbf{85}, 022502 (2012).

\bibitem{Hiyama:2005cf}
E.~Hiyama, M.~Kamimura, A.~Hosaka, H.~Toki and M.~Yahiro,
Phys. Lett. B \textbf{633}, 237-244 (2006).

\bibitem{Zhao:2016ccp}
J.~Zhao, H.~He and P.~Zhuang,
Phys. Lett. B \textbf{771}, 349-353 (2017).

\bibitem{Yang:2018oqd}
G.~Yang, J.~Ping and J.~Segovia,
Phys. Rev. D \textbf{99}, no.1, 014035 (2019).

\bibitem{Wang:2019rdo}
G.~J.~Wang, L.~Meng and S.~L.~Zhu,
Phys. Rev. D \textbf{100}, no.9, 096013 (2019).

\bibitem{Crater:1987hm} 
  H.~W.~Crater and P.~Van Alstine,
  Phys.\ Rev.\ D {\bf 36}, 3007 (1987).
  
\bibitem{Crater:2010fc} 
  H.~W.~Crater and J.~Schiermeyer,
  Phys.\ Rev.\ D {\bf 82}, 094020 (2010).  

\bibitem{Shi:2013rga} 
  S.~Shi, X.~Guo and P.~Zhuang,
  Phys.\ Rev.\ D {\bf 88}, no. 1, 014021 (2013).  

\bibitem{Tiesinga:1993zza}
E.~Tiesinga, B.~J.~Verhaar and H.~T.~C.~Stoof,
Phys. Rev. A \textbf{47}, 4114-4122 (1993).

\bibitem{Alford:2013jva} 
  J.~Alford and M.~Strickland,
  Phys.\ Rev.\ D {\bf 88}, 105017 (2013).  

\bibitem{Guo:2015nsa} 
  X.~Guo, S.~Shi, N.~Xu, Z.~Xu and P.~Zhuang,
  Phys.\ Lett.\ B {\bf 751}, 215 (2015).
  

\bibitem{Chen:2020xsr}
S.~Chen, J.~Zhao and P.~Zhuang,
Phys. Rev. C \textbf{103}, no.3, L031902 (2021).

\bibitem{grimm1971monte}
R.~C. Grimm and R.~G. Storer, J. Comput. Phys. {\bf 7}, 134 (1971).

\bibitem{foulkes2001quantum}
M. Foulkes, L. Mitas, R. Needs, G. Rajagopal, Rev. Mod. Phys. {\bf 73}, 33 (2001).

\bibitem{feagin2002quantum}
J.M. Feagin, {\it Quantum Methods with Mathematica}, Springer, New York (1994).

\bibitem{roy2001time}
A.K. Roy, N. Gupta and B.M. Deb, Phys. Rev. {\bf A 65}, 012109 (2001).

\bibitem{wadehra2003ground}
A. Wadehra, A.K. Roy and B.M. Deb, Int. J. Quantum Chem. {\bf 91}, 597 (2003).

\bibitem{sullivan2005determining}
D.M. Sullivan and D.S. Citrin, J. Appl. Phys. {\bf 97}, 104305 (2005).

\bibitem{sullivan2005determining2}
D.M. Sullivan, J. Appl. Phys. {\bf 98}, 084311 (2005).

\bibitem{roy2005low}
A.K. Roy, A.J. Thakkar, and B.M. Deb, J. Phys. A: Math. Gen. {\bf 38}, 2189 (2005).

\bibitem{Sudiarta:2007}
I.~W.~Sudiarta and D.~J.~W. Geldart, J. Phys. A: Math. Theor. {\bf 40}, 1885 (2007).

\bibitem{Strickland:2009ft}
M.~Strickland and D.~Yager-Elorriaga,
J. Comput. Phys. \textbf{229}, 6015-6026 (2010).

\bibitem{1994JCoPh.115..470C}
   H.~W.~Crater,
   J. Comput. Phys.~{\bf 115}, 470 (1994).

\bibitem{code}
   Source code of the current paper is publicly available at the following Github repository [\href{https://github.com/ShuzheShi/inverse-power-method-schroedinger-solver}{https://github.com/ShuzheShi/inverse-power-method-schroedinger-solver}].
\end{thebibliography}
\end{document}